\documentclass[prl,twocolumn,showpacs,preprintnumbers,amsmath,amssymb,
showkeys]{revtex4}

\usepackage{graphicx}
\usepackage{dcolumn}
\usepackage{bm}
\begin{document}

\preprint{APS/123-QED}

\title{Optimal branching asymmetry of hydrodynamic pulsatile trees}

\author{Magali Florens}
\email{magali.florens@cmla.ens-cachan.fr}
\affiliation{CMLA, ENS Cachan, CNRS, UniverSud, 61 Avenue du Pr\'esident Wilson,
F-94230 Cachan}
\author{Bernard Sapoval}
\author{Marcel Filoche}
\affiliation{Laboratoire de Physique de la Mati\`{e}re Condens\'{e}e,
CNRS, Ecole Polytechnique, 91128 Palaiseau, France}

\date{\today}

\begin{abstract}
  Most of the studies on optimal transport are done for steady state
  regime conditions. Yet, there exists numerous examples in living
  systems where supply tree networks have to deliver products in a
  limited time due to the pulsatile character of the flow. This is the
  case for mammals respiration for which air has to reach the gas
  exchange units before the start of expiration. We report here that
  introducing a systematic branching asymmetry allows to reduce the
  average delivery time of the products. It simultaneously increases
  its robustness against the unevitable variability of sizes related
  to morphogenesis. We then apply this approach to the human
  tracheobronchial tree. We show that in this case all extremities are
  supplied with fresh air, provided that the asymmetry is smaller than
  a critical threshold which happens to fit with the asymmetry
  measured in the human lung. This could indicate that the structure
  is adjusted at the maximum asymmetry level that allows to feed all
  terminal units with fresh air.
\end{abstract}



\maketitle

Branched transportation networks are ubiquitous in living
systems. Such trees have been suggested to allow efficient feeding
the body volume from a small source, the aorta for blood or the mouth
for oxygen~\cite{Weibel1963,West1997,Mauroy2004,Morel2008,Mauroy2010,
Dodds2010,Corson2010}. For example, the human tracheobronchial tree
brings fresh air to the oxygen-blood exchange units, called the {\em
acini}, that fill the majority of the volume of the thoracic
cage~\cite{Weibel1963}.

In the general perspective of how statistical physics may help to
better understand the relation between structure and physiological
function, we focus here on pulsatile trees in which the delivery of
products has to be achieved in a limited time. For example in mammals
respiration, the respiratory cycle is made on two successive steps,
inspiration and expiration. The period of this cycle is about 5
seconds for humans at rest (2~s for inspiration, 3~s for expiration).
And of course, the transit time from the mouth to the acini has to be
short enough so that expiration does not start before the arrival of
fresh air into the acini. The present work analyses the possible
statistical constraints related to this last condition.

From the point of view of ventilation, the conducting airway system
can be modelled as an arrangement of pipes defined by their diameter
and length. The branchings are essentially dichotomous, each airway
being divided into two smaller daughter airways. A branching defines
the beginning of a new generation. The tracheobronchial tree starts at
the trachea (generation~0) and ends in the terminal bronchioles
(around generation~15) at the entrance of the acini~\cite{Weibel1963}.
Since no gas exchanges take place in the tracheobronchial tree, its 
volume is referred to as the {\em dead space volume} (DSV). Its value is
around 170~mL in the human lung~\cite{Weibel1984}. To better understand
human respiration in relation with the tree structure, we discuss
successively the general properties of two models, respectively
symmetric and asymmetric, before discussing the real human airway
system.

The first model has symmetrical branching. Although the human
airway system is both non symmetric and exhibits some type of
randomness, a basic step in describing its morphology has been the
introduction of the so called Weibel's ``A'' model~\cite{Weibel1963}.
In this first model, the tree is likened to a hierarchical network of
cylindrical pipes with symmetrical branching and a uniform scaling
ratio $h_0 = 2^{-1/3} \approx 0.79$ between the airway sizes of
consecutive generations. This value corresponds to the classical
Murray-Hess law~\cite{Hess1914,Murray1926} for which the diameter of
the mother branch $d_0$ and the diameters of both daughter branches
$d_1$ and $d_2$ are linked by the relationship $d_0^3 = d_1^3 +
d_2^3$.

An important second step in modeling the human lung morphology has
been the introduction from anatomical studies of a systematic
asymmetry~\cite{Horsfield1971,Raabe1976,Phillips1997,Majumdar2005}. This
is the second model studied in this paper. The branching asymmetry is
characterized by two different scaling ratios, $h_{0,max}=0.88$ and
$h_{0,min}=0.68$~\cite{Majumdar2005}. Each parent airway gives rise to
a larger daughter airway (the major airway) and a smaller daughter
airway (the minor airway). Note that $(0.88)^3 + (0.68)^3$ = 1 so that
the asymmetric tree respects the above dissipation requirement. And
so, the Weibel's A model is the exact symmetrisation of the real
asymmetric structure.

Those are the two models to be compared, including the possible role
of a statistical noise associated to anatomical variability. Note that
two different types of structural randomness can appear in these
models: the first source of disorder, due to the systematic branching
asymmetry, is found in the random succession of large and small
airways along any given airway path. The second type of disorder (the
variability) results from the statistical noise of the biological
growth process~\cite{Weibel1984} and spreads the distribution of the
branch sizes at a given generation even in the symmetric model.

The criterion that will be used in the following to quantify the
delivery performance of the tree is the distribution of oxygenation
times of fresh air into the acini. The oxygenation time in one acinus
is obtained by substracting from the total duration of the inspiratory
phase, $t_{ins}$, the time spent in the extrathoracic airways,
$t_{ext}$ (approximately constant and equal to 0.47~s at
rest~\cite{Sandeau2010}), and the transit time from the trachea to
this acinus, $t_{tr}$: $$t_{ox} = t_{ins} - t_{ext} - t_{tr}$$

The performances of symmetric and of asymmetric trees have been
computed on 15~generations trees. The geometrical parameters used in
our computations, are summarized in Table~\ref{tab:param}. These are
the scaling ratios at each generation and the length to diameter ratio
for each generation. It has to be noted that the values of the dead
space volumes are kept almost similar. Due to the uniform motion of
the thoracic cage, each acinus is assumed to act as an hydrodynamic
pump draining the same flux. In other terms, the gas exchange units
are equitably ventilated~\cite{Weibel2005}. As a consequence, starting
from the bottom of the tree, any two daughters of a given mother
branch create, independently of their sizes, the same additive flux in
their mother branch. The time spent in a branch is then directly
obtained from the flux and the branch size. In the inspiratory phase
studied here, the total flow can be considered approximately
constant in time with a velocity in the trachea of about 1~m/s
~\cite{Weibel1984}. Since the duration of the inspiration $t_{ins}$
and the time spent in the extrathoracic airways $t_{ext}$ are the same
for all acini, the oxygenation times in the acini are thus entirely
determined by the transit times of fresh air from the trachea to the
terminal bronchioles.
\begin{table}[htbp]
\caption{Model parameters}
\label{tab:param}
\begin{ruledtabular}
\begin{tabular}{c|c|c|c}
  Model& Scaling ratio for $D$ &Ratio $L/D$ \footnote{D and L:
  diameter and length of the airway.}&DSV (mL) \\ \hline Symmetric &
  $h_0=2^{-1/3}$ & 3.00 & 220 \\ \hline Asymmetric & $h_{0,min}=0.68$
  & 3.00 & 213\\ & $h_{0,max}=0.88$ & &\\ \end{tabular}
\end{ruledtabular}
\end{table}
\begin{figure}[htbp]
  \centering
  \includegraphics[width=0.3\textwidth]{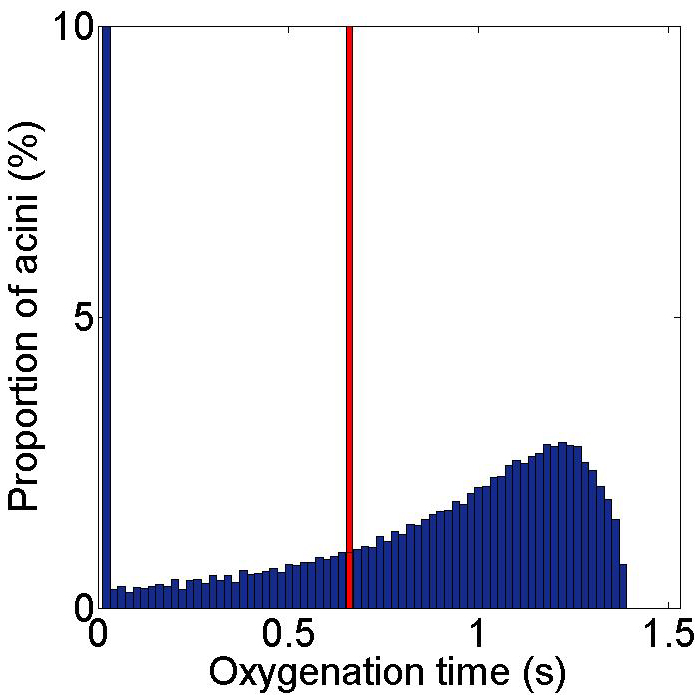}
  \caption{Distribution of oxygenation times in the acini for both
  symmetric (red) and asymmetric (blue) trees. All the pathways have
  the same transit times in the symmetric case, hence a single peak
  distribution of oxygenation times at 0.67~s. In the asymmetric
  case the distribution is spread with a mean
  value about 0.82~s and a standard deviation of 0.43~s.}
\label{fig:distrib0}
\end{figure}

Fig.~\ref{fig:distrib0} shows the distribution of the acini
oxygenation times for both models. The distribution for the symmetric
tree trivially presents a single peak at $t=0.67$~s as all the pathways
from trachea to a terminal bronchiole are identical. The distribution
of oxygenation times for the asymmetric tree is spread around a mean
value of 0.82~s and has a standard deviation of 0.43~s. Therefore, on
average, fresh air arrives slighty sooner and remains longer in acini
supplied by an asymmetric tracheobronchial tree than by a symmetric
one. (Such an asymmetric tree structure can be described as
multifractal~\cite{Zamir2001}).

We now study how the performances of these tree structures are robust
or not with respect to anatomical variability. To mimic this
variability, we introduce a ``growth noise'' by adding Gaussian
variations of the scaling ratios. At each branching, the values of the
scaling ratios are modified to become random variables:
\begin{equation}
\begin{cases}
  h_{min} = h_{0,min} \left( 1 + \sigma X \right)\\
  h_{max} = h_{0,max} \left( 1 - \sigma X \right)
\end{cases}
\label{eq:disorder}
\end{equation}
$X$ being a centered gaussian random variable of standard deviation 1.
The mean values $h_{0,min}$ and $h_{0,max}$ of these random variables
correspond to the values given in table~\ref{tab:param}. Scaling
ratios of different bifurcations are assumed to be independent random
variables while in the same branching, scaling ratios $h_{min}$ and
$h_{max}$ are anticorrelated. This means that if the random variable
for $h_{max}$ take a larger (resp. smaller) value than its mean value,
then $h_{min}$ is very likely to take a smaller (resp. larger) value
than its mean value.
\begin{figure}[htbp]
  \centering
  \includegraphics[width=0.3\textwidth]{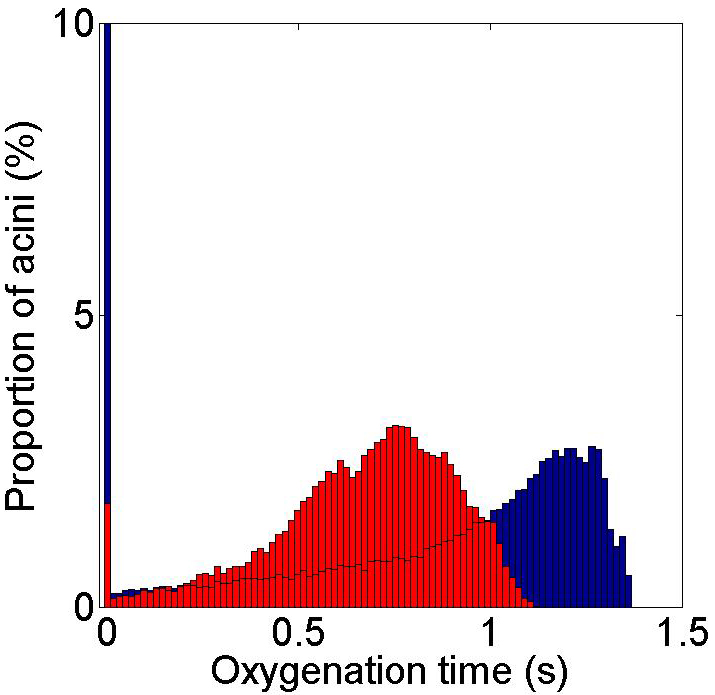}
  \caption{Distribution of oxygenation times with randomized scaling
  ratios according to Eq.~\ref{eq:disorder}: symmetric tree in red and
  asymmetric tree in blue. The oxygenation times are spread in the
  symmetric case, but the distribution is shifted towards smaller
  oxygenation times. The distribution for the asymmetric tree is not
  modified by randomization.}
\label{fig:distrib0_noise}
\end{figure}

The results are shown in Fig.~\ref{fig:distrib0_noise} which displays
the distributions of oxygenation times for both symmetric and
asymmetric trees with randomized scaling ratios ($\sigma = 5\%$). The
distribution for the symmetric tree is now also spread, with an
average oxygenation time of 0.67~s and a standard deviation of 0.28~s.
Unlike the symmetric case, the distribution of times in the asymmetric
model is not modified by the randomization of the scaling ratios (mean
value of 0.82~s and standard deviation of 0.44~s). The delivery of
products through a dichotomous tree of constant depth thus appears to
be more efficient both for average oxygenation time and immunity
versus growth fluctuations.

So far, the two model trees that we have considered had an equal
number of generations for all pathways. In fact, the real human
tracheobronchial tree is even more complex because not only the
branchings are asymmetric but also the number of generations is not
uniform~\cite{Horsfield1971,Raabe1976}. The human tracheobronchial
tree does not end at a constant generation but at a constant airway
diameter, that of the terminal bronchioles, around
$0.5$~mm~\cite{Weibel1984,Weibel2005}. The terminal generations of the
tree range from 8 to 22~\cite{Horsfield1971} and the length of the
first generations airways also exhibits specific features linked to
anatomical constraints ~\cite{Weibel1963,Raabe1976, Phillips1995}. The
diameter scaling ratios keep equal to $0.68$ and $0.88$ for all
generations. The aspect ratios (length over diameter) are specific for
the first 4 generations (respectively equal to 3.07, 1.75, 1.43, 
and 1.85) and equal to 3.00 for higher generations (generations 6 to 22).
The distribution of oxygenation times for the real human tree 
(not shown here) is found to have an average oxygenation time
of 0.67~s and a standard deviation of 0.13~s.

The asymmetry level, namely $h_{0,max}=0.88$ and $h_{0,min}=0.68$,
used in the above computations was considered as a given fact drawn
from anatomical measurements. It is a natural question to ask for a
reason of such values. For obvious reasons, the branching asymmetry
cannot be too strong because it would lead to a structure with only
very few wide pathways surrounded by a large number of much narrower
pathways with large hydrodynamic resistances. Moreover, due to flux
conservation, the wider pathways correspond to larger transit
times. If the asymmetry level were too important, a number of
extremities would not be supplied with fresh air because the transit
time in their pathway would be too long. We will now investigate the
influence of the asymmetry level on the acini oxygenation. The
question that naturally arises is thus the following: how much
asymmetry can there be? In other words, can one define an optimal
asymmetry level?

\begin{figure}[htbp]
\centering \includegraphics[width=0.40\textwidth]{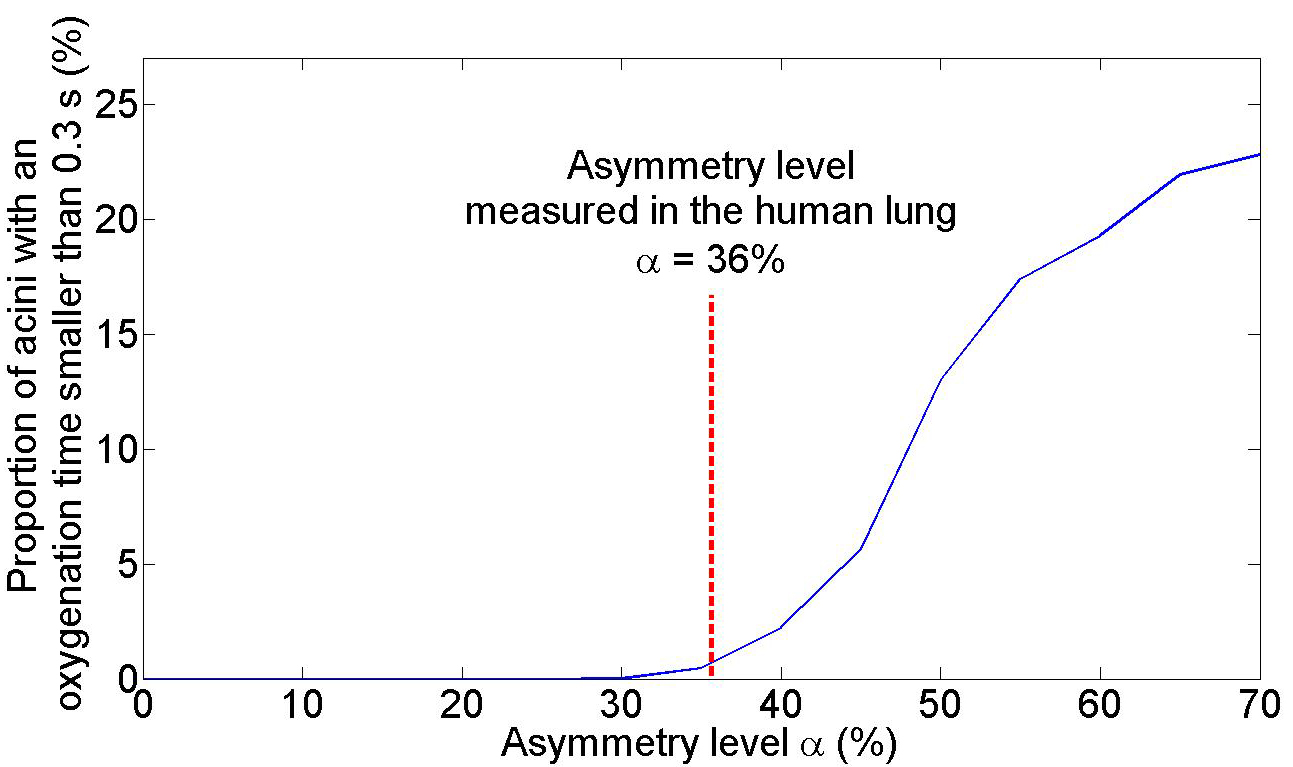}
\caption{Proportion of unactive acini (with oxygenation time smaller than
  0.3~s) as a function of the asymmetry level $\alpha$. All acini are
  found to be active provided that the asymmetry level is below
  $\alpha \approx 35\%$. This value almost exactly corresponds to the
  measured asymmetry level in the human lung (36\%).}
\label{fig:fresh0}
\end{figure}

In order to investigate this question, trees of different asymmetry
levels have been studied. In this study, the asymmetry level is
characterized by one parameter $\alpha$ such that:
\begin{equation}
\begin{cases}
\displaystyle h_{0,max}^3 = h_0^3 \left(1+\alpha\right) \\
\displaystyle h_{0,min}^3 = h_0^3 \left(1-\alpha\right)
\end{cases}
\label{eq:alpha}
\end{equation}
The scaling ratios 0.88 and 0.68 as measured in the human
tracheobronchial tree would for instance correspond to an asymmetry
level of 36\% ($\alpha=0.36$). All computed trees are conditionned to
have the same thoracic volume (dead space volume + acini volume) and
the aspects ratios $L/D$ given above. Considering that fresh air has
to remain at least 0.3~s in the acinar region in order to achieve the
gas exchange process (a duration consistent with computations of the
dynamical diffusion oxygen transport in the
acinus~\cite{Filoche2008}), we have computed for each asymmetry level
the proportion of acini with an oxygenation time larger than 0.3~s. This
corresponds to a total transit time from the mouth to the terminal
bronchiole smaller than 1.7~s. Results are presented in
Fig.~\ref{fig:fresh0}. The proportion of acini fed with fresh air for
more than 0.3~s is 100\% for the symmetric tree (zero asymmetry level)
and remains 100\% until a threshold value of the asymmetry level. As
one can see on Fig.~\ref{fig:fresh0}, this threshold value is about
35\%, almost identical to the value measured in the human
tracheobronchial tree.

\begin{figure}[htbp]
  \centering
  \includegraphics[width=0.37\textwidth]{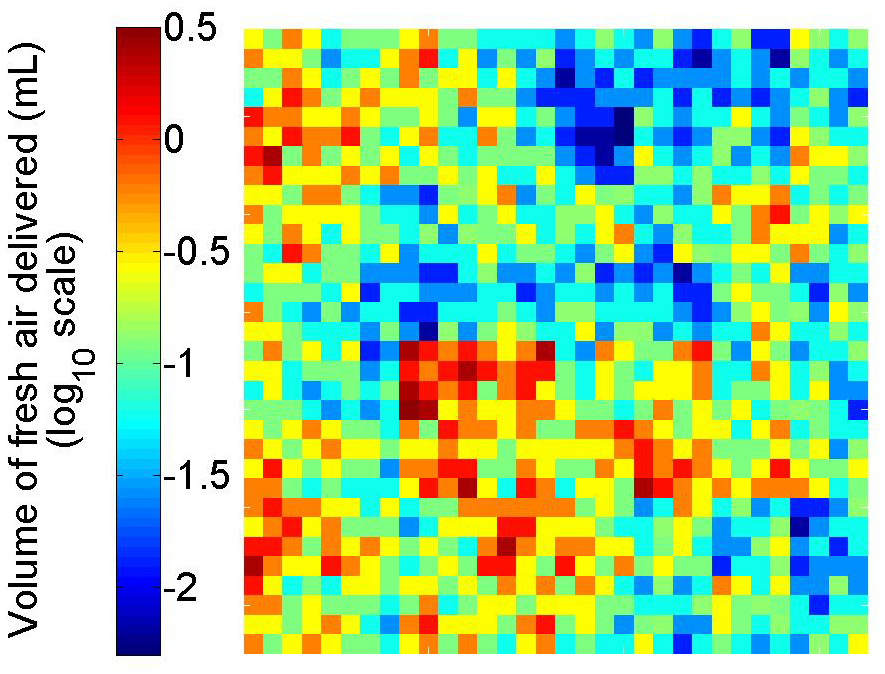}
  \caption{2D representation of the distribution of volumes of fresh
  air delivered at generation 10 in the human tracheobronchial
  tree (parameters described in the text).}
  \label{fig:2d}
\end{figure}

So it seems that the systematic asymmetry found in the
tracheobronchial tree corresponds to the maximum value that allows to
feed all acini. The advantages of such a structure are several: the
number of acini with fresh air is 100\%, and the distribution of the
ventilated volumes is robust against anatomical variability. Note that
the spread distribution of arrival times in the exchange units may
contribute to smoothen the oxygen delivery to the blood. On the other
hand, due to this spread distribution of transit times, all acini do
not receive the same volume of fresh air. Even if all acini receive
the same flux, the uneveness of branching creates an inhomogeneity of
the volumes of supplied fresh air that multiplicatively increases at
each generation. In the limit of an infinite tree, this would
mathematically lead to a multifractal distribution of the volume of
fresh air delivered in the extremities~\cite{Sapoval1997}. This
pre-multifractal behavior can be observed in Fig.~\ref{fig:2d} which
shows a 2D~representation of the distribution of volumes of fresh air
delivered at generation 10. One observes a wide spread of the
distribution of fresh air volumes. This could indicate that a
dynamical regulation of airway diameters might be necessary to
minimize such effects, as already suggested to fight inertial effects
in the upper part of the airway tree~\cite{Mauroy2003}. Note that the
spread in the distribution of external gas in the acini would also
induce an inherent noise in NMR imaging of the lung.

In summary, numerical computations show that symmetric branching trees
are not optimal to supply a volume when a constraint of limited
delivery time is imposed. For trees of uniform depth, the average
transit time is found to be smaller when the branching is asymmetric
rather than symmetric. Moreover, the distribution of oxygenation times
in the acini of the asymmetric tree is almost not modified by a
stochastic variability of the sizes of the branches. This approach is
then extended to the study of the role of variable asymmetry in trees,
with the same trachea, the same ratio length over diameter, the same
diameter of the terminal bronchioles and the same inner volume. It is
shown then that there exists a maximum asymmetry level above which the
number of terminal units, or acini, supplied with fresh air, departs
from 100\%. Interestingly, this maximum value corresponds almost
exactly to the asymmetry level measured in the human lung. The
geometry of the lung airways thus appears as if being adjusted to have
the largest possible branching asymmetry, while still being able to
feed efficiently all acini with fresh air. It may be considered as
remarkable that an asymmetric tree works better and that the natural
selection of mammalian seems to have found a level of asymmetry that
can be considered as best from the physical point of view.

\begin{acknowledgments}
The authors would like to thank Pr E.R.~Weibel for fruitful
discussions.
\end{acknowledgments}

\bibliography{Bibliography}

\end{document}